\documentclass[preprint,12pt]{elsarticle}

\usepackage{amssymb}
\usepackage{color}

\journal{JCAP}

\begin{document}

\begin{frontmatter}



\title{There is no universal acceleration scale in galaxies}


\author{Man Ho Chan$^1$, Shantanu Desai$^2$ \& Antonino Del Popolo$^{3,4}$}

\address{$^1$ Department of Science and Environmental Studies, The Education University of Hong Kong, Tai Po, New Territories, Hong Kong, China \\
$^2$ Department of Physics, Indian Institute of Technology, Hyderabad, Telangana-502285, India \\
$^3$ Dipartimento di Fisica e Astronomia, University of Catania, Viale Andrea Doria 6, 95125, Catania, Italy \\
$^4$ Institute of Astronomy, Russian Academy of Sciences, Pyatnitskaya str. 48, 119017 Moscow, Russia}

\ead{chanmh@eduhk.hk}

\begin{abstract}
Recently, many studies seem to reveal the existence of some correlations between dark matter and baryonic matter. In particular, the unexpected tight Radial Acceleration Relation (RAR) discovered in rotating galaxies has caught much attention. The RAR suggests the existence of a universal and fundamental acceleration scale in galaxies, which seems to challenge the $\Lambda$CDM model and favour some modified gravity theories. A large debate about whether RAR is compatible with the $\Lambda$CDM model has arisen. Here, by analysing the high quality velocity dispersion profiles of 13 E0-type elliptical galaxies in the SDSS-IV MaNGA sample and assuming a power-law function of radius $r$ for the 3-dimensional velocity dispersion in each galaxy, we report the RAR for E0-type elliptical galaxies and we show that the resultant RAR has more than $5\sigma$ deviations from the RAR in late-type galaxies. This new RAR provides an independent probe to falsify the existence of any universal acceleration scale in galaxies. Our result significantly challenges those modified gravity theories which suggest the existence of any universal acceleration scale.
\end{abstract}


\end{frontmatter}



\section{Introduction}
Dark matter is one of the most profound mysteries in astrophysics. Many observations provide indirect hints for the existence of dark matter, while there is no signal obtained from direct detections \citep{Tan,Aprile,Abecrcrombie} and indirect detections \citep{Desai,Ackermann,Albert,Chan4,Bergstrom,Cavasonza,Chan5}. The standard cosmological model (the $\Lambda$CDM model) has many successful descriptions in the large-scale structures of the universe (e.g. the power spectrum and the cosmic microwave background) \citep{Croft,Planck}. However, the $\Lambda$CDM model seems to have some discrepancies between observations and predictions on small scales, such as the missing satellite problem \citep{Moore}, the core-cusp problem \citep{deBlok}, the too-big-to-fail problem \citep{Boylan}, the satellite planes problem \citep{Pawlowski}, and the problem of the slope and scatters in the baryonic Tully-Fisher relation \citep{McGaugh4} (see the review in \citet{Bullock} for the small-scale problem in the $\Lambda$CDM model). Also, the $\Lambda$CDM model predicts that the interaction between dark matter and baryons is nearly negligible so that there should be no strong correlation between dark matter and baryons (e.g. the separation between hot gas and dark matter in the Bullet Cluster). Nevertheless, many recent studies of galaxies and galaxy clusters have revealed some strong correlations between dark matter and baryons \cite{McGaugh,McGaugh2,Lelli,Chan,Chan2}. 

In particular, there is a tight relation connecting the dynamical radial acceleration (mainly affected by dark matter) and the baryonic radial acceleration (the acceleration component contributed by baryonic mass) in spiral galaxies, which is called the Radial Acceleration Relation (RAR) \citep{McGaugh2,McGaugh3}. The empirical form of the RAR is written as \citep{McGaugh2}
\begin{equation}
g_{\rm dyn}=\frac{g_{\rm bar}}{1-e^{-\sqrt{g_{\rm bar}/a_0}}},
\end{equation}
where $g_{\rm dyn}$ and $g_{\rm bar}$ are the dynamical radial acceleration and the baryonic radial acceleration respectively. This relation is obtained by analysing the data of 153 late-type galaxies in the SPARC sample \citep{McGaugh2}. The RAR obtained from the SPARC data (hereafter called sRAR) suggests the existence of a fundamental acceleration scale of $a_0= (1.20 \pm 0.26) \times 10^{-10}$ m/s$^2$ \citep{McGaugh2}. The small scatters in the sRAR seem to reveal that it is tantamount to a natural law for rotating galaxies.

Moreover, another study adding 25 early-type galaxies (12 elliptical and 13 lenticular) and 62 dwarf spheroidal galaxies to the SPARC sample gives a new composite RAR, which is very similar to the sRAR \citep{Lelli2}. This result suggests that the alleged fundamental acceleration scale may be universal for all galaxies with different morphologies, although only a small percentage of early-type galaxies ($\approx 10$\%) is involved.

Generally speaking, the existence of the acceleration scale is not obvious for the $\Lambda$CDM model, though a few studies have reproduced the RAR using it \citep{Desmond,Ludlow,Stone,Paranjape}. These imply that the RAR could be an emergent phenomenon in the $\Lambda$CDM model. Nevertheless, many recent studies have pointed out that the existence of the universal acceleration scale is better interpreted as a possible sign of modified gravity \citep{Milgrom2,Li,Green}. For example, Modified Newtonian Dynamics (MOND) and Emergent Gravity have predicted the existence of a universal acceleration scale \citep{Milgrom,Verlinde}. Therefore, the existence of a fundamental and universal acceleration scale has become a very hot issue in astrophysics and cosmology now.

However, there is still an on-going debate on the existence of such fundamental acceleration scales in galaxies \citep{McGaugh3,Rodrigues}. Also, if the fundamental acceleration scale originates from modified gravity, it must be universal and valid for elliptical galaxies and galaxy clusters as well. In particular, some recent studies show that the RAR in elliptical galaxies is consistent with the sRAR \citep{Lelli2,Chae} while some other studies show that the RAR in galaxy clusters have large scatters and does not agree with the sRAR \citep{Tian,Chan3,Pradyumna,Pradyumna2}. Some studies also reveal that the sRAR is not obeyed for dwarf disk spirals and low-surface-brightness galaxies \cite{Paolo}. Therefore, the existence of a fundamental acceleration scale in galaxies and the universality of the acceleration scale is controversial. No smoking gun evidence has been obtained to confirm or falsify the above claims. In this article, we report the RAR for E0-type elliptical galaxies and we show that the resultant RAR has more than $5\sigma$ deviations from the RAR in late-type galaxies. Therefore, this new RAR can provide an independent probe and a strong evidence to falsify the existence of any universal acceleration scale in galaxies.

\section{Velocity dispersion profiles of the E0-type elliptical galaxies}
Recently, some high-quality observations of elliptical galaxies have been obtained by galaxy surveys, such as MaNGA \citep{Aguado} and ATLAS$^{\rm 3D}$ \citep{Cappellari}. We follow the selected galaxies in the MaNGA sample in \cite{Chae} to produce the RAR for the elliptical galaxies (hereafter called eRAR). There are 15 E0-type elliptical galaxies chosen from the MaNGA sample in \cite{Chae} based on the following criteria: 1. E0-type with small probability of being type-S0, 2. bulge-dominates light, 3. ellipticity $<0.1$, 4. S\'ersic index $>3$, 5. slow rotators, and 6. enough independent values of velocity dispersion to construct a velocity dispersion radial profile. On top of these criteria in \cite{Chae}, we finally discard 2 more galaxies in our analysis because they don't have enough data points to do the Markov Chain Monte Carlo (MCMC) analysis. Therefore, we will analyze 13 galaxies in this study.

The E0-type elliptical galaxies chosen are all slow rotators, which are completely different from the rotating galaxies analysed in the SPARC sample. Therefore, analysing the E0-type elliptical galaxies can be regarded as an independent probe to test the existence of the universal acceleration scale. Besides, the almost spherical structure can give smaller systematic uncertainties related to anisotropy or irregularities so that spherical symmetry can be assumed throughout the analysis. 

The idea of using E0-type elliptical galaxies to do the RAR analysis is not new. Previous studies have used them to examine the universality of the RAR \citep{Chae,Rong}. However, to calculate $g_{\rm dyn}$, these studies have either assumed the MOND interpolating function \citep{Chae} or some universal functional form of the total (dark matter) mass density profile \citep{Rong}, which are quite model-dependent. These assumptions would involve a systematic bias, because using any presumed universal functional form of $g_{\rm dyn}$ might generate some potential connections with $g_{\rm bar}$ and give lower scatters. In our study, we do not assume any universal form of dark matter density profile. Instead, we assume that the 3-D velocity dispersion profile of each galaxy can be described by a power-law function (see below) and apply the Jeans equation, a fundamental equation in gravitation, to calculate the values of $g_{\rm dyn}$. Following the Jeans equation only to derive $g_{\rm dyn}$ can ensure our analysis to be more model-independent and contain fewer systematic uncertainties.

We first transform the line-of-sight velocity dispersion maps obtained in MaNGA \citep{Aguado} into the line-of-sight velocity dispersion profile $\sigma_{\rm los}(R)$ for each chosen galaxy, where $R$ is the projected radius. We take the azimuthal averaging of the velocity dispersion in concentric bins for different angular radii from the galactic centre. The angular distance for each bin is taken as 0.5". The small fluctuations of the velocity dispersion in each bin are represented by the $1\sigma$ standard deviation. Here, we neglect all the data within the region of $R \le 2"$ due to the point spread function size of the observational data. It is because the observational uncertainties would be large within twice the finite fiber size of the MaNGA observations ($\approx 1"$). We therefore discard the data within a diameter of 2". To ensure the high quality of the analysis, we also neglect the data with signal-to-noise ratio $S/N<10$, which are mainly found in the large $R$ region. After neglecting the two extreme regions, we only analyze the data within the reliable region of the projected radius for each galaxy to minimise the uncertainties (see Fig.~1 for the line-of-sight velocity dispersion profile for the 9047-6102 galaxy). Generally speaking, the line-of-sight velocity dispersion $\sigma_{\rm los}(R)$ can be well described by a power-law function $\sigma_{\rm los}(R) \propto R^{-\gamma}$ for all 13 chosen galaxies (see Fig.~2 for the other 12 galaxies). The reduced $\chi^2$ values ($\chi_{\rm red}^2$) are all smaller than 0.3 (see Table 1), which mean very good fits have been obtained for the power-law function \footnote{Note that the error bars of the line-of-sight velocity dispersion for each galaxy are obtained from raw observational data, which are independent of the power-law fits. The reduced $\chi^2$ values indicate the quality of fits for the power-law function. Generally speaking, good fits are obtained if the reduced $\chi^2$ values are smaller than 1.}.

Note that the data reduction process in our analysis is a little bit different from the one in \cite{Chae}. The study in \cite{Chae} has defined the radial bins with an equal number of velocity dispersions at spaxels (constant number of raw data approach) while we have defined the radial bins with constant angular difference (0.5") (constant-bin approach). Since we have followed the power-law function of $R$ in our fitting procedures, we need to guarantee the smoothness of the line-of-sight velocity dispersion profiles. The overall fits would be more sensitive to the radial dependence. This is the reason why we adopt the constant-bin approach in the data reduction process. The smoothness of the velocity dispersion profiles can better justify our assumption of using the 3-D velocity dispersion power-law function. Moreover, the error bars in the velocity dispersion profiles depend on the distribution of the observed raw data. If there are too many raw data concentrated at a certain $R$, our approach might generate larger error bars and a smaller reliable region. However, if the raw data density is too small for some $R$, the approach used in \cite{Chae} would generate a large uncertainty in $R$ because they need to combine the raw data for a large range of $R$ to obtain one velocity dispersion data point. A certain wide gaps in $R$ might appear and affect the smoothness of the velocity dispersion profile. This also generates different uncertainties in $R$ in the profile. Nevertheless, the intrinsically smooth and continuous MOND-interpolating function is assumed in the analysis of \cite{Chae} so that they need not care about the smoothness problem in their analysis. Since our analysis and the study in \cite{Chae} are based on different assumptions, there are slight differences in the data reduction process. Both approaches would have advantages and disadvantages. Nevertheless, both approaches are considering the same amount of information stored in the observed raw data.  

\section{The theoretical framework}
In the standard theoretical framework, the line-of-sight velocity dispersion profile of a galaxy is given by \citep{Binney,Agnello}
\begin{equation}
\sigma_{\rm los}^2(R)=\frac{2}{\Sigma(R)} \int_R^{\infty} \left(1-\beta \frac{R^2}{r^2} \right) \frac{\sigma_r^2(r) \rho(r)}{\sqrt{r^2-R^2}}rdr,
\end{equation}
where $\Sigma(R)$ is the surface stellar mass density profile, $\beta=1-\sigma_t^2/\sigma_r^2$ is the anisotropy parameter, $\sigma_t$ is the 3-D tangential velocity dispersion, $\sigma_r(r)$ is the 3-D radial velocity dispersion and $\rho(r)$ is the 3-D stellar mass density profile. The surface stellar mass density profile can be written as $\Sigma(R)=I(R)\Upsilon(R)$, where $\Upsilon(R)=\Upsilon_0 \times \max \{ 1+K[2.33-3(R/R_e)],1 \}$ is the mass-to-light ratio profile and $I(R)$ is the surface brightness profile \citep{Chae}. Here, $\Upsilon_0$ is a constant, $R_e$ is the effective radius of a galaxy and $K$ is a constant parameter. The surface brightness profile of an elliptical galaxy can be best described by the S\'ersic profile \citep{Cote}:
\begin{equation}
I(R)=I_0e^{-b_n[(R/R_e)^{1/n}-1]},
\end{equation}
where $b_n=1.9992n-0.2371$ and $n$ is the S\'ersic index. The values of $R_e$ and $n$ for each galaxy can be fitted independently by the observed surface brightness profile \citep{Chae}. Therefore, we need not fit these values by the observed velocity dispersion data. The values of $R_e$ and $n$ for each galaxy are shown in Table 2. We maximise the likelihood between the eRAR and sRAR to get the product of $I_0 \times \Upsilon_0$. Here, the likelihood is proportional to $e^{-\chi^2}$ so that minimising $\chi^2$ would give a maximised likelihood. The value of $\chi^2$ will be defined later. Moreover, the 3-D stellar mass density profile can be written in terms of the surface stellar mass density profile as
\begin{equation}
\rho(r)=-\frac{1}{\pi} \int_r^{\infty} \frac{d \Sigma(R)}{dR} \frac{dR}{\sqrt{R^2-r^2}}.
\end{equation}
For the anisotropy parameter $\beta$, we follow the benchmark model as \cite{Chae}
\begin{equation}
\beta(r)=\beta_0+(\beta_1-\beta_0) \frac{(r/r_a)^2}{1+(r/r_a)^2},
\end{equation}
where $\beta_0$, $\beta_1$ and $r_a$ are constant parameters. In fact, the type-E0 elliptical galaxies are all slow-rotators, which have a relatively small rotation velocity component. On the other hand, to calculate $\sigma_{\rm los}(R)$, we need to assume a certain form of $\sigma_r(r)$. Since observational data suggest a power-law form of $\sigma_{\rm los}(R) \propto R^{-\gamma}$, we also take the power-law function $\sigma_r(r)=\sigma_0(r/r_a)^{-\alpha}$ to model the 3-D velocity dispersion $\sigma_r(r)$ for each galaxy, where $\sigma_0$ and $\alpha$ are free parameters for fitting. 

Note that $\sigma_{\rm los}(R)$ is a projected quantity such that any bumps or kinks in $\sigma_{\rm los}(R)$ would probably give a drastic change in $\sigma_r(r)$. Nevertheless, most of the observed $\sigma_{\rm los}(R)$ can be well-fitted by a smooth power-law form within the uncertainties. Moreover, it can be shown that the conversion between $\sigma_r(r)$ and $\sigma_{\rm los}(R)$ would almost maintain the power-law form when the anisotropy parameter $\beta$ is small in magnitude \citep{Wolf}. Therefore, assuming a power-law form of $\sigma_r(r)$ would be a very good approximation and avoid creating too many free parameters. In fact, assuming a power-law form of $\sigma_r(r)$ is the key difference with respect to previous work, like the study in \cite{Chae}. Here, taking the power-law form of $\sigma_r(r)$ is based on the observed power-law form of $\sigma_{\rm los}(R)$, which would give a more justified option compared with the MOND interpolating function used in \cite{Chae} or mass parametrized model used in \cite{Rong}. It is because many studies have revealed a large variety of mass functions in galaxies (e.g. Burkert profile, Navarro-Frenk-White profile, pseudo-isothermal profile, isothermal profile) \cite{Spano,Naray,Velander}, which could not be simply described by a single function or parametrized model. As a result, based on this assumption, altogether $\sigma_{\rm los}(R)$ depends on six independent parameters $\{K, \beta_0, \beta_1, r_a, \sigma_0, \alpha \}$ for each galaxy. 

To fit the observed velocity dispersion profile $\sigma_{\rm los}(R)$ for each chosen galaxy, we perform the MCMC analysis, using the {\tt emcee}~\citep{emcee} sampler, to get the best-fit values of the six independent parameters $\{K, \beta_0, \beta_1, r_a, \sigma_0, \alpha \}$. Originally, there are 15 MaNGA E0-type elliptical galaxies selected in \cite{Chae} to perform the analysis. However, there are two galaxies (8718-3704 and 9088-3701) in which the total number of data points for $\sigma_{\rm los}(R)$ are less than 6 (i.e. the number of fitted parameters). Therefore, we neglect these two galaxies and we only focus on 13 E0-type elliptical galaxies. We have input the following priors for the parameters: $K=[0,1]$, $\beta_0=[-2,+0.7]$, $\beta_1=[-2,+0.7]$, $r_a=[0.1R_e,R_e]$, $\sigma_0=[50,400]$ km/s, $\alpha=[-1,+1]$. We use the same ranges of priors in \citep{Chae} for $K$, $\beta_0$, $\beta_1$ and $r_a$. For $\sigma_0$ and $\alpha$, the chosen ranges of the priors are based on the observed line-of-sight velocity dispersion profiles ($\sigma_{\rm los} \sim 200$ km/s and $\alpha \sim 0$). After performing the MCMC analysis, we get the best-fit values of the parameters and their $1\sigma$ ranges for each galaxy (see Table 3). Although some recent studies have pointed out that using the MCMC analysis may give absurd results due to the choice of priors \citep{Li2}, our fitted parameters are within reasonable ranges and the priors assumed are all physically motivated. Moreover, to avoid any systematic bias, using flat priors would be better than using any Gaussian prior as we need not presume any unjustified prior central values used in the Gaussian prior.

The six fitted parameters can determine the values of the dynamical radial acceleration $g_{\rm dyn}$ as a function of the spherical radius $r$. The enclosed dynamical mass can be obtained by the Jeans equation \citep{Wolf}:
\begin{equation}
g_{\rm dyn}=\frac{\sigma_r^2}{r}(\gamma_*+\gamma_{\sigma}-2\beta).
\end{equation}
where $\gamma_*=-d \ln \rho/d \ln r$ and $\gamma_{\sigma}=-d \ln \sigma_r^2/d \ln r=2 \alpha$. 

Using the six parameters fitted from the MCMC analysis, we can generate $g_{\rm dyn}$ at different $r$ for each galaxy. For the baryonic radial acceleration, we have
\begin{equation}
g_{\rm bar}=\frac{GM_{\rm bar}}{r^2}=\frac{G}{r^2} \int_0^r4 \pi r'^2\rho(r')dr'.
\end{equation}
The value of the baryonic radial acceleration $g_{\rm bar}(r)$ slightly depends on the value of $K$ and the unknown product $I_0 \times \Upsilon_0$. However, the possible range of $I_0 \times \Upsilon_0$ depends on the stellar population, which is quite uncertain. Here, we adopt the value of $I_0 \times \Upsilon_0$ such that it could maximise the likelihood between the resultant eRAR and the sRAR. We first get $g_{\rm dyn}(r)$ by the MCMC analysis for each galaxy. Then we write $g_{\rm bar}(r)=I_0\Upsilon_0g(r)$, where $g(r)$ is a function calculated by Eq.~(7). Therefore, for each galaxy, we can have the functional form of the eRAR $g_{\rm dyn}(g_{\rm bar}(I_0\Upsilon_0))$ and we compare this functional form with the one in Eq.~(1) (i.e. the sRAR). By minimising the $\chi^2$ between the eRAR and sRAR, we can get the value of $I_0\times \Upsilon_0$ for each galaxy. Consequently, such an optimization would give $g_{\rm dyn} \approx g_{\rm bar}$ in the large $g_{\rm bar}$ regime. In other words, the value of $I_0 \times \Upsilon_0$ would be the maximised value. This is justified because the enclosed total mass profiles in most galaxies are dominated by baryonic components at their central regions (i.e. large $g_{\rm bar}$). Following these steps, we can compute the resultant eRAR ($g_{\rm dyn}$ vs. $g_{\rm bar}$). 

In our MCMC analysis, we got the best-fit values of the parameters $\{K, \beta_0, \beta_1, r_a, \sigma_0, \alpha \}$ and their $1\sigma$ limits by more than 1000 iterations (see Fig.~3 and Fig.~4 for the corner plots of two galaxies, and Table 1 for the $\chi^2$ values of the MCMC analysis). We have calculated the auto-correlation time for two galaxies to check the convergence. We have also cross-checked that the best-fit parameters remain almost the same after 5000 iterations. These indicate that the MCMC chains have converged in our analysis. Note that the observed $\sigma_{\rm los}$ is not very sensitive to the values of $K$, $\beta_0$, $\beta_1$ and $r_a$. Therefore, the best-fit values are less clear so that the allowed ranges would be relatively larger. Here, the values of $K$ obtained in this analysis are generally consistent with that in previous studies $K \sim 0.53^{+0.05}_{-0.04}$ for other similar galaxies \cite{Chae3}. Besides, the average values of $\beta$ tend to be negative for many galaxies in our analysis. Previous studies show that the distribution of the average $\beta$ based on the Monte Carlo sets of models of the ATLAS$^{\rm 3D}$ galaxies is more on the side of negative values (see Fig.~21 of \cite{Chae3}). This is also the reason why the input prior ranges of $\beta_0$ and $\beta_1$ are negative skewed ($[-2,+0.7]$). Therefore, the values generated in our analysis are reasonable.

Using the best-fit values, we can get the `best-fit eRAR' (the BEST scenario) for each galaxy. We also consider the $1\sigma$ upper and lower limits of the parameters, which can give two extreme sets of $g_{\rm dyn}$ correspondingly (the MAX and MIN scenarios). Although all these three scenarios will be considered in our analysis, the BEST scenario (the more likely scenario) is regarded as the benchmark for our consideration and comparison. Note that the six parameters might not be completely independent so that some intrinsic correlations might exist among them. Such potential correlations would decrease the actual uncertainties in the MAX and MIN scenarios. Therefore, since we have not considered these potential correlations, our MAX and MIN scenarios would give more conservative ranges of $g_{\rm dyn}$ in our analysis. Moreover, the values of $g_{\rm bar}$ is insensitive to the uncertainty of $K$. The other dependence such as the S\'ersic index $n$ and the effective radius $R_e$ can be determined by independent observations of the galactic surface brightness profiles. Therefore, the uncertainty of $g_{\rm bar}$ is nearly negligible in the $\log g_{\rm bar}$ space.  

\section{Data analysis}
Firstly, we perform the `individual galaxy analysis'. We take the null hypothesis as follows: the eRAR of an individual galaxy is identical to the sRAR represented by Eq.~(1). Therefore, we compare the likelihood between the eRAR of each galaxy for three different scenarios and the sRAR. The likelihood can be characterized by the $\chi^2$ value, which is defined as 
\begin{equation}
\chi^2=\sum_i \frac{(g_{{\rm dyn},i}-g_{\rm dyn,sRAR})^2}{\sigma_{\rm sRAR}^2}
\end{equation}
where $g_{{\rm dyn},i}$ are the dynamical radial acceleration calculated in Eq.~(6) for different $r$ (i.e. different $g_{{\rm bar},i}$), $g_{\rm dyn,sRAR}$ are the dynamical radial acceleration for different $g_{{\rm bar},i}$ predicted by Eq.~(1), and $\sigma_{\rm sRAR}$ is the $1 \sigma$ variation of the $g_{\rm dyn,sRAR}$ in Eq.~(1). The $1\sigma$ variation is $g_{{\rm bar},i}$-dependent and it originates from the uncertainty of $a_0$. Although the values of $g_{{\rm bar},i}$ come from the modeling, they are the independent variables of the sRAR in Eq.~(1). Generally speaking, we are comparing the likelihood between the eRAR based on our modeling and the empirical form of the sRAR. Such a comparison depends on the relation $g_{\rm dyn}(g_{{\rm bar},i})$ between two models, but not the input values of $g_{{\rm bar},i}$. Therefore, the model-dependency of $g_{{\rm bar},i}$ would not be a problem here. The value of $\chi^2$ for each galaxy can determine the $p-$value of the null hypothesis. 

Generally speaking, among the three scenarios, the MAX scenario gives the worst fits while the MIN scenario gives the best fits for all galaxies. Following the BEST scenario, the null hypothesis is rejected at $5\sigma$ (the probability of the null hypothesis being true is $p<5 \times 10^{-7}$) for 11 out of 13 galaxies (see Table 2). For the MIN scenario, the null hypothesis is ruled out at $5\sigma$ for 6 galaxies. Therefore, based on the `individual galaxy analysis', the eRAR is very unlikely to be identical to the sRAR (i.e. RAR is unlikely to be universal). Moreover, we perform the fits of $a_0$ by using Eq.~(1) for each galaxy. Following the BEST scenario, the best-fit $a_0$ can range from $0.2 \times 10^{-10}$ m/s$^2$ to $4.2 \times 10^{-10}$ m/s$^2$ (see Table 2). Combining the $a_0$ values for the 13 galaxies gives $a_0=(2.0 \pm 1.1)\times 10^{-10}$ m/s$^2$. The large range of $a_0$ implies that the existence of a fundamental acceleration scale for the E0-type elliptical galaxies is quite unlikely.

Apart from the `individual galaxy analysis', we also combine all $(g_{\rm bar}, g_{\rm dyn})$ data together for the 13 galaxies in the BEST scenario to perform a more holistic stacked analysis. We first sort the data and then smooth the data by averaging 10 points of $(g_{\rm bar}, g_{\rm dyn})$ in a bin. We can get an overall eRAR for the 13 galaxies (see Fig.~5). We can see that a significantly large deviation between the eRAR and sRAR (the red line) can be seen when $g_{\rm bar}$ is small. In fact, since we have maximised the likelihood between the eRAR and the sRAR (by choosing the appropriate values of $\Upsilon_0$), the deviation between them in the large $g_{\rm bar}$ regime ($\log g_{\rm bar}>-9$) is very small. For the sRAR, the observational data give $g_{\rm dyn} \approx g_{\rm bar}$ for $\log g_{\rm bar} \ge -9.12$ \citep{McGaugh2}. Therefore, to avoid systematic bias, we mainly focus on comparing the eRAR and sRAR for $\log g_{\rm bar}<-9.3$ (outside the $2\sigma$ deviation of $g_{\rm bar}$ from $\log g_{\rm bar}=-9.12$). In this region, the likelihood between the eRAR and the sRAR is extremely small. The probability is $p<3.5\times 10^{-9}$ for the null hypothesis being true (i.e. ruled out at more than $5.9 \sigma$). From Fig.~5, we can see that the sRAR basically lies outside the $5\sigma$ boundary of the eRAR for the entire region $\log g_{\rm bar}<-9.3$. On the other hand, we could maximise the likelihood between the eRAR and the sRAR in the small acceleration regime. However, the whole eRAR in Fig.~5 would be shifted downwards and it would give a very large deviation in the high acceleration regime. The overall deviation would be much larger. Therefore, the RAR and the alleged fundamental acceleration scale is unlikely to be universal. 

\section{Discussion}
In fact, as we have mentioned, previous studies \cite{Chae,Rong} have analysed the type-E0 galaxies and examined the universality of the RAR. These studies claim that the resultant eRAR is consistent with the sRAR while our study does not agree with this claim. The major reason is that these studies have assumed either the MOND interpolating function \citep{Chae} or a universal functional form of dark matter mass density profile (the Navarro-Frenk-White profile) \citep{Rong}, which might generate intrinsic systematic bias in the analysis. Strictly speaking, these assumptions are not necessary and may not be correct. Although another previous study has shown that the simple MOND interpolating function is preferred by the elliptical galaxies in the ATLAS$^{\rm 3D}$ sample \citep{Chae2}, it does not entail that the simple MOND interpolating function used in \cite{Chae} is also consistent with the sample of MaNGA. Even though we can use a generalised form of the interpolating function with an extra parameter \citep{Chae2}, the possible forms connecting $g_{\rm dyn}$ and $g_{\rm bar}$ would not be exhaustive. There might not exist any universal functional form of $g_{\rm dyn}(g_{\rm bar})$ for elliptical galaxies in general. In fact, one recent study using the data of the Kilo-Degree Survey (KiDS-1000) has shown that the RAR for early-type galaxies is significantly different (more than $6\sigma$ deviation) from the RAR for late-type galaxies \citep{Brouwer}. This may support our choice of not assuming any functional form of $g_{\rm dyn}(g_{\rm bar})$ prior to the analysis. Furthermore, our results are generally consistent with the one in \cite{Brouwer}.

In our study, we have mainly followed the Jeans equation to perform the analysis. The Jeans equation is a fundamental equation in gravitation, which is model-independent and reliable. The resultant dark matter density profile for each galaxy may not follow any specific functional form, which can reveal the real behavior of the eRAR. Nevertheless, we have taken two specific functional forms in modeling the anisotropy parameter $\beta$ and the 3-D radial velocity dispersion $\sigma_r(r)$. For the functional form of $\beta$, since the magnitude of $\beta$ is small ($|\beta|<1$), the effect of $\beta$ on $g_{\rm dyn}$ is subdominant as the empirical values of $\gamma_*$ are larger than 2. Moreover, the effect of the radial dependent term $(\beta_1-\beta_0)\{(r/r_a)^2/[1+(r/r_a)^2]\}$ on $\beta$ is relatively small so that the values of $\beta$ mainly depends on the parameter $\beta_0$. Therefore, the systematic uncertainty contributed by the functional form of $\beta$ used in our analysis is very small. For the functional form of $\sigma_r(r)$, we have taken a power-law function because the observed line-of-sight velocity dispersion profile $\sigma_{\rm los}(R)$ can be best described by a power-law function for each galaxy. Besides, as the values of $\alpha$ are all very small ($\alpha \sim 0$, see Table 3), the values of $g_{\rm dyn}$ would be mainly determined by the values of $\sigma_0$ rather than the power-law factor $(r/r_a)^{-\alpha}$. Therefore, the systematic uncertainty contributed due to the functional forms of $\sigma_r(r)$ used is also very small. 

Furthermore, we have compared the eRAR with the alleged universal functional form of the RAR (i.e. Eq.~(1)), but not the actual distribution of the sRAR (i.e. $(g_{\rm bar}, g_{\rm dyn})$). In fact, our objective is not going to compare the distributions of the $(g_{\rm bar}, g_{\rm dyn})$ between elliptical galaxies and spiral galaxies. We aim at showing that the acceleration scale $a_0$ might not be a universal constant for both elliptical and spiral galaxies. Since the acceleration scale $a_0$ is primarily defined under Eq.~(1) (with the smallest possible scatters) \cite{McGaugh2}, we finally compare our distribution of the $(g_{\rm bar}, g_{\rm dyn})$ in elliptical galaxies with the functional form of Eq.~(1).

In our analysis, only 13 type-E0 elliptical galaxies are involved and they are of the same morphology. In principle, as they are almost spherical in shape and belong to the same class, only relatively little variations exist among these galaxies. Also, the E0-type elliptical galaxies (slow-rotators) and the fast rotating spiral galaxies belong to two completely different extremes in terms of morphology and galactic dynamics. Therefore, these properties ensure that the RAR obtained from the E0-type elliptical galaxies can provide an independent probe to test the alleged universal nature of the RAR, derived mainly by the rotating galaxies. Although a previous study of the RAR has included 25 early-type galaxies in the sample \citep{Lelli2}, the percentage of elliptical galaxies is low and only 1 of them is a type-E0 galaxy. In view of this, using a RAR derived purely from type-E0 galaxies can give a critical test for the claimed universal RAR. Although only a small number of a single morphological type of galaxies are considered in this analysis, it is already enough to falsify the existence of the universal acceleration scale in galaxies. This result is also consistent with some previous findings analysing galaxy clusters \citep{Tian,Chan3,Pradyumna}. As a result, it would challenge those modified gravity theories which suggest the existence of a universal acceleration scale (e.g. MOND and Emergent Gravity).

\begin{figure}
\vskip15mm
 \includegraphics[width=140mm]{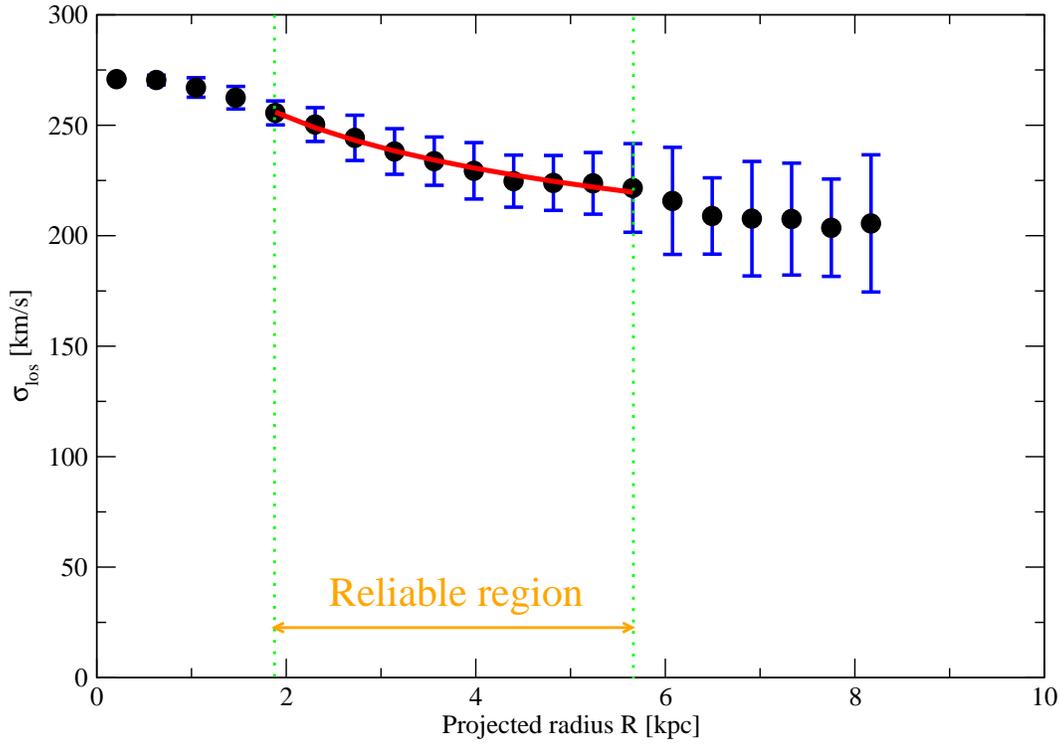}
 \caption{The line-of-sight velocity dispersion profile of the 9047-6102 galaxy. We neglect the data at $R \le 2"$ ($R \le 1.9$ kpc) and the data with signal-to-noise ratio $<10$ ($R \ge 5.7$ kpc). The red solid line represents the best-fit power-law form $\sigma_{\rm los}=280~{\rm km/s}(R/\rm 1~kpc)^{-0.14}$ for the data in the reliable region.}
\vskip5mm
\end{figure}

\begin{figure}
\vskip15mm
 \includegraphics[width=140mm]{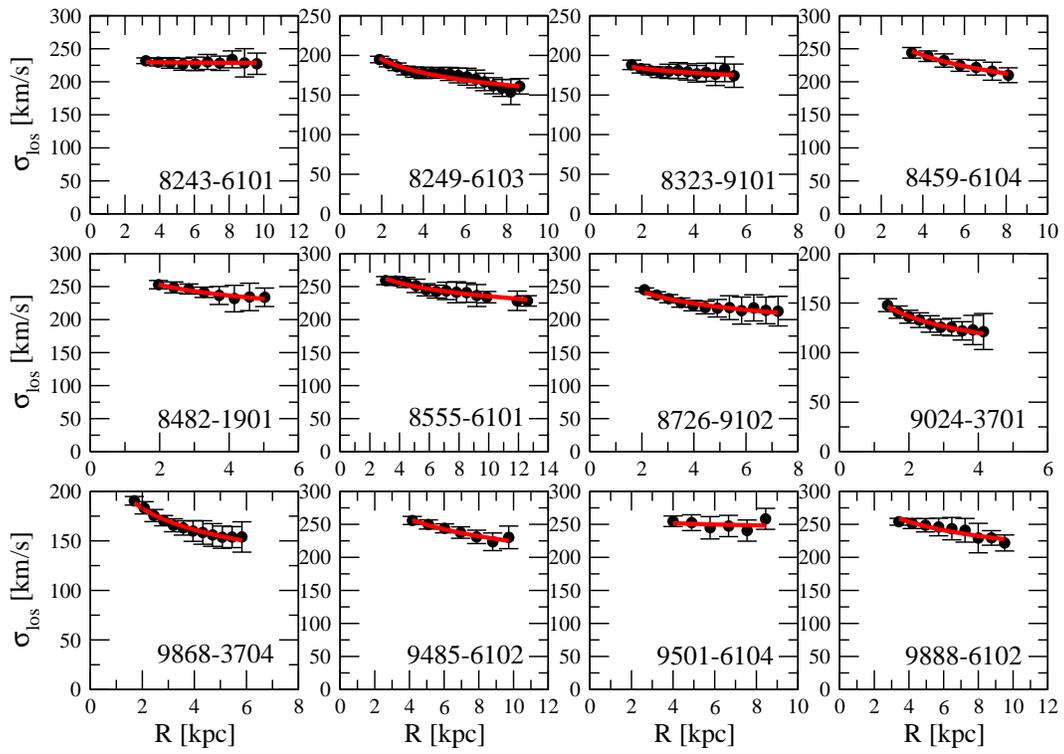}
 \caption{The line-of-sight velocity dispersion profiles of the other 12 chosen galaxies within the reliable regions (the velocity dispersion profile of the 9047-6102 galaxy is shown in Fig.~1). The red solid lines represent the best-fit power-law functions for the data.}
\vskip5mm
\end{figure}

\begin{figure}
\vskip15mm
 \includegraphics[width=140mm]{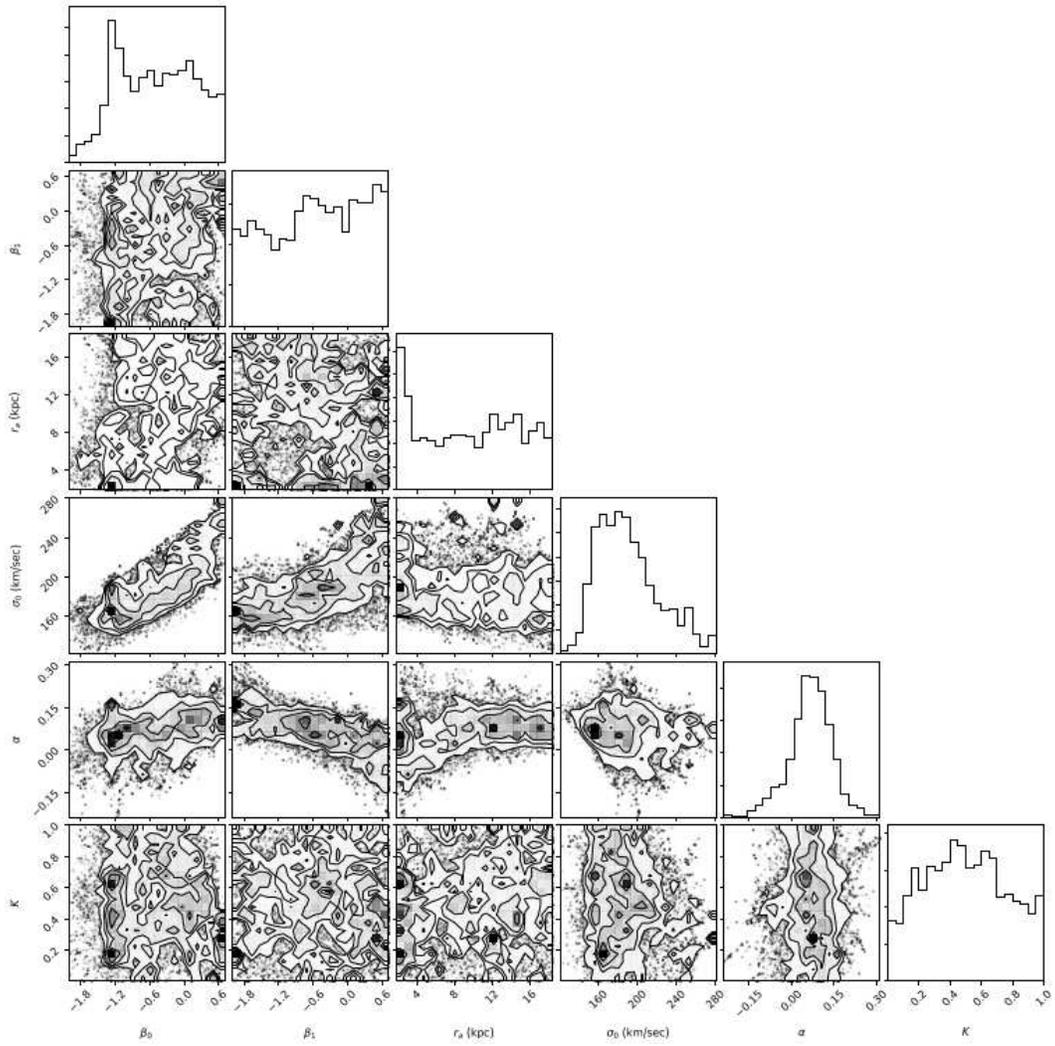}
 \caption{The corner plot of the 6 parameters for the 8726-9102 galaxy.}
\vskip5mm
\end{figure}

\begin{figure}
\vskip15mm
 \includegraphics[width=140mm]{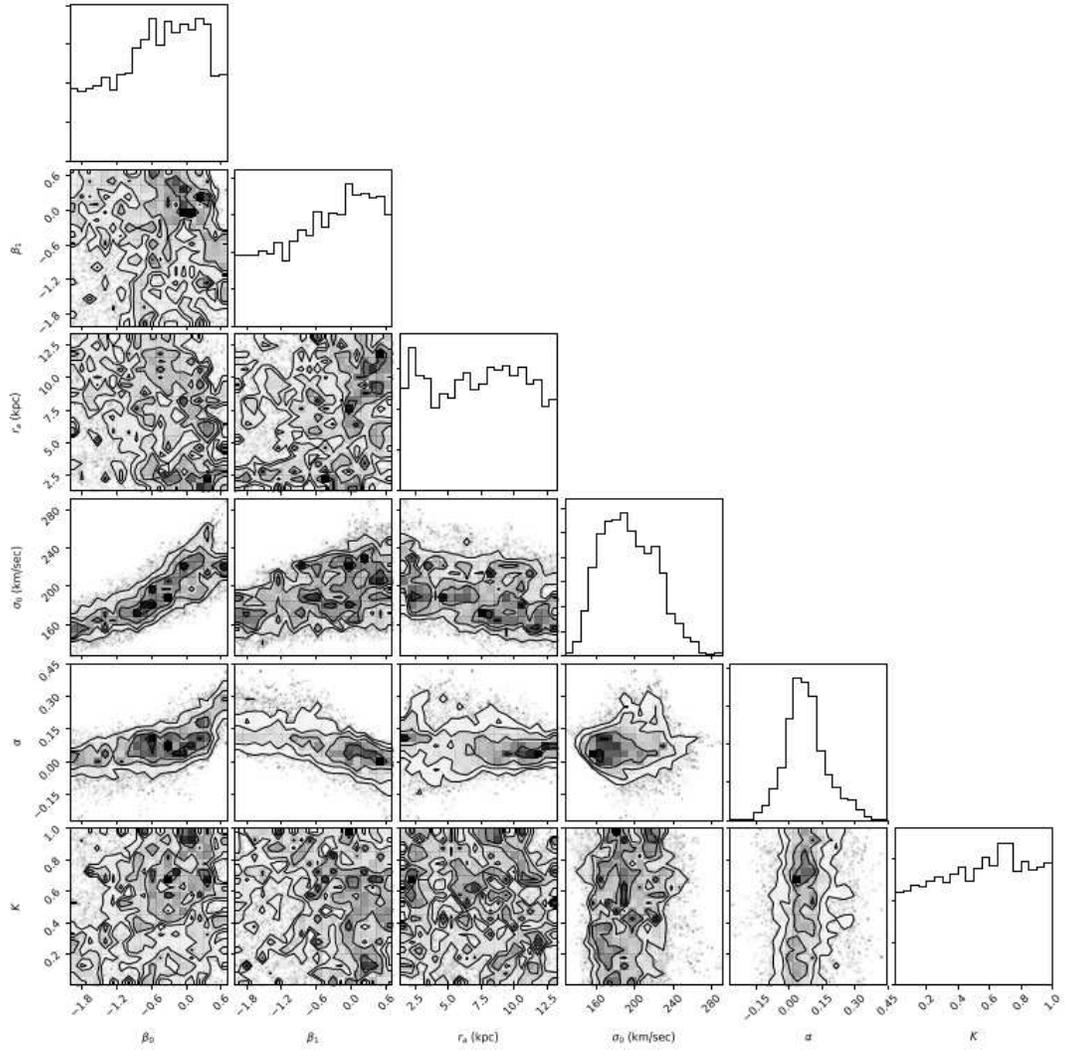}
 \caption{The corner plot of the 6 parameters for the 9047-6102 galaxy.}
\vskip5mm
\end{figure}

\begin{figure}
\vskip10mm
 \includegraphics[width=140mm]{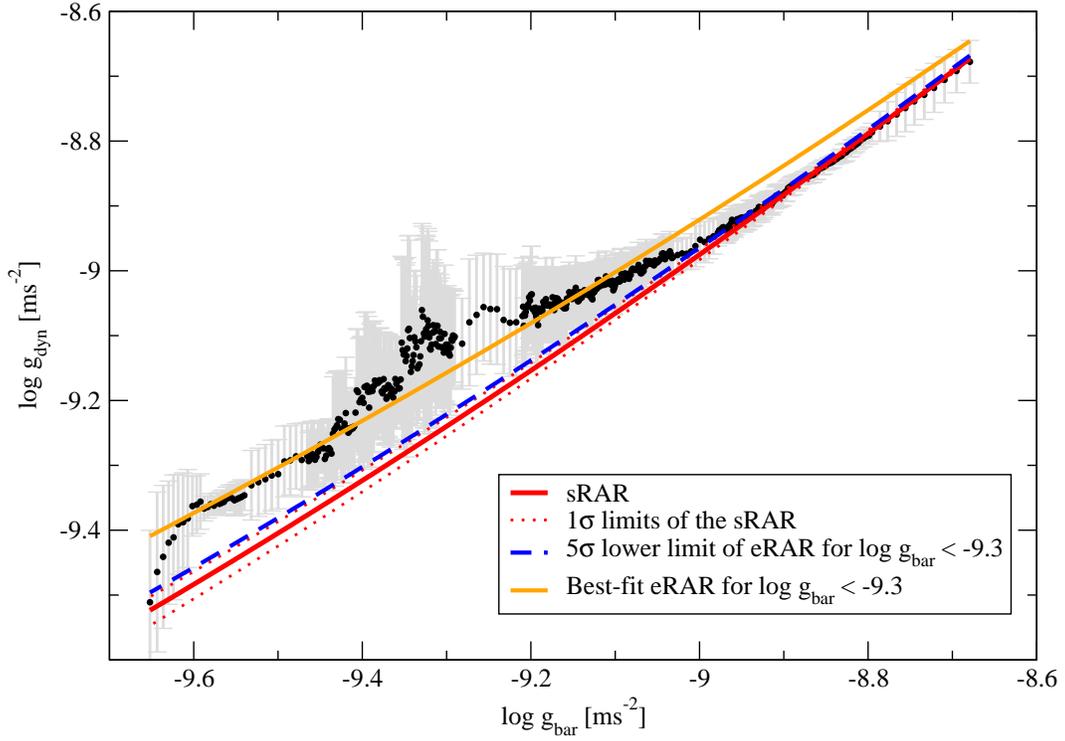}
 \caption{The black dots and grey error bars represent the eRAR derived from the 13 E0-type elliptical galaxies (the BEST scenario). The orange solid line represents the best-fit eRAR following Eq.~(1) (best-fit $a_0=3.1\times 10^{-10}$ m/s$^2$). The blue dashed line indicates the $5\sigma$ lower limit of the eRAR following Eq.~(1). The red solid and dotted lines are the sRAR and its $1\sigma$ limits respectively.}
\vskip5mm
\end{figure}
 

\begin{center}
\begin{table}
\caption{The reduced $\chi^2$ value ($\chi_{\rm red}^2$) of the power-law fits of $\sigma_{\rm los}(R)$, and the $\chi^2$ values and the degrees of freedom (dof) obtained from the MCMC analysis.}
 \label{table2}
 \begin{tabular}{@{}lccc}
  \hline
  Galaxy & $\chi_{\rm red}^2$ (power-law) & $\chi^2$ (MCMC) & dof (MCMC) \\
  \hline
  \hline
  8243-6101 & 0.07 & 9.38 & 4 \\
  8249-6103 & 0.18 & 9.96 & 11  \\
  8323-9101 & 0.07 & 2.46 & 6 \\
  8459-6104 & 0.03 & 3.14 & 1  \\
  8482-1901 & 0.03 & 2.77 & 2  \\
  8555-6101 & 0.06 & 0.99 & 7  \\
  8726-9102 & 0.28 & 5.72 & 6  \\
  9024-3701 & 0.02 & 1.22 & 4  \\
  9047-6102 & 0.02 & 0.87 & 4  \\
  9485-6102 & 0.07 & 2.80 & 1  \\
  9501-6104 & 0.23 & 1.04 & 0  \\
  9868-3704 & 0.07 & 1.36 & 6  \\
  9888-6102 & 0.21 & 2.44 & 3  \\
  \hline
 \end{tabular}
\end{table}
\end{center}

\begin{center}
\begin{table}
\caption{The S\'ersic index $n$ \citep{Chae}, the effective radius $R_e$ \citep{Chae}, the probability ($p$-value) of the null hypothesis being true for the MIN and BEST scenarios, and the best-fit value of $a_0$ for each galaxy. For $p<5 \times 10^{-7}$, the null hypothesis is rejected at $5\sigma$.}
 \label{table3}
 \begin{tabular}{@{}lccccc}
  \hline
  Galaxy & $n$ & $R_e$ (kpc) & $p$-value (MIN) & $p$-value (BEST) & best-fit $a_0$ ($10^{-10}$ m/s$^2$) \\
  \hline
  \hline
  8243-6101 & 6.411 & 21.16 & $0.45$ & $1.3\times 10^{-6}$ & $1.4$  \\
  8249-6103 & 5.288 & 9.17 &  $<5\times 10^{-7}$ & $<5\times 10^{-7}$ & $1.7$ \\
  8323-9101 & 7.369 & 14.14 & $0.17$ & $<5\times 10^{-7}$ & $1.4$ \\
  8459-6104 & 7.245 & 10.43 & $<5\times 10^{-7}$ & $<5\times 10^{-7}$ & $0.2$ \\
  8482-1901 & 8 & 24.17 & $0.99$ & $0.47$ & $1.2$ \\
  8555-6101 & 4.873 & 11.7 & $<5\times 10^{-7}$ & $<5\times 10^{-7}$ & $3.3$ \\
  8726-9102 & 8 & 18.4 & $0.99$ & $<5\times 10^{-7}$ & $1.4$ \\
  9024-3701 & 6.003 & 4.69 & $0.08$ & $<5\times 10^{-7}$ & $1.7$ \\
  9047-6102 & 6.965 & 13.3 & $0.98$ & $<5\times 10^{-7}$ & $1.4$ \\
  9485-6102 & 5.498 & 12.8 & $<5\times 10^{-7}$ & $<5\times 10^{-7}$ & $3.8$ \\
  9501-6104 & 5.08 & 15.76 & $0.72$ & $<5\times 10^{-7}$ & $1.5$ \\
  9868-3704 & 5.997 & 6.46 & $<5\times 10^{-7}$ & $<5\times 10^{-7}$ & $3.2$  \\
  9888-6102 & 4 & 9.82 & $<5\times 10^{-7}$ & $<5\times 10^{-7}$ & $4.2$ \\
  \hline
 \end{tabular}
\end{table}
\end{center}

\begin{center}
\begin{table}
\caption{The 6 best-fit parameters obtained from the MCMC analysis.}
 \label{table1}
 \begin{tabular}{@{}lcccccc}
  \hline
  Galaxy & $K$ & $\beta_0$ & $\beta_1$ & $r_a$ (kpc) & $\sigma_0$ (km/s) & $\alpha$ \\
  \hline
  \hline
  8243-6101 & $0.479^{+0.084}_{-0.163}$ & $-0.295^{+0.218}_{-0.402}$ & $-0.429^{+0.445}_{-0.791}$ & $11.04^{+1.82}_{-1.94}$ & $193.93^{+18.50}_{-12.40}$ & $0.032^{+0.108}_{-0.084}$ \\
  8249-6103 & $0.510^{+0.339}_{-0.334}$ & $-0.571^{+0.826}_{-0.876}$ & $-0.870^{+1.036}_{-0.768}$ & $4.98^{+2.62}_{-2.52}$ & $144.26^{+16.23}_{-16.12}$ & $0.088^{+0.125}_{-0.124}$ \\
  8323-9101 & $0.584^{+0.327}_{-0.383}$ & $-0.075^{+0.618}_{-1.548}$ & $-0.514^{+0.863}_{-0.929}$ & $4.34^{+5.31}_{-1.73}$ & $158.58^{+36.68}_{-28.75}$ & $-0.050^{+0.138}_{-0.086}$ \\
  8459-6104 & $0.517^{+0.334}_{-0.346}$ & $-0.642^{+0.919}_{-0.849}$ & $-0.711^{+0.922}_{-0.897}$ & $4.92^{+3.23}_{-2.71}$ & $197.06^{+32.34}_{-25.22}$ & $0.17^{+0.12}_{-0.14}$ \\
  8482-1901 & $0.395^{+0.315}_{-0.271}$ & $-0.681^{+0.809}_{-0.774}$ & $-0.890^{+1.053}_{-0.723}$ & $14.19^{+6.78}_{-9.90}$ & $192.85^{+44.48}_{-23.77}$ & $0.020^{+0.103}_{-0.080}$ \\
  8555-6101 & $0.496^{+0.292}_{-0.324}$ & $-0.594^{+0.562}_{-0.775}$ & $-0.553^{+0.783}_{-1.033}$ & $6.17^{+3.80}_{-3.65}$ & $206.64^{+36.09}_{-22.61}$ & $0.063^{+0.090}_{-0.085}$ \\
  8726-9102 & $0.488^{+0.306}_{-0.287}$ & $-0.497^{+0.747}_{-0.726}$ & $0.520^{+0.864}_{-0.990}$ & $9.82^{+5.73}_{-6.70}$ & $185.34^{+38.18}_{-26.08}$ & $0.072^{+0.063}_{-0.071}$ \\
  9024-3701 & $0.393^{+0.402}_{-0.285}$ & $-0.368^{+0.704}_{-0.803}$ & $-0.421^{+0.618}_{-0.879}$ & $2.60^{+1.32}_{-1.54}$ & $103.70^{+16.85}_{-12.84}$ & $0.030^{+0.142}_{-0.129}$ \\
  9047-6102 & $0.543^{+0.319}_{-0.340}$ & $-0.074^{+0.501}_{-0.590}$ & $-0.269^{+0.660}_{-1.008}$ & $6.14^{+4.95}_{-2.93}$ & $209.39^{+27.37}_{-28.94}$ & $0.084^{+0.130}_{-0.084}$ \\
  9485-6102 & $0.530^{+0.288}_{-0.346}$ & $-0.276^{+0.689}_{-1.160}$ & $-0.910^{+0.953}_{-0.692}$ & $6.82^{+3.87}_{-3.32}$ & $201.50^{+39.44}_{-21.85}$ & $0.166^{+0.168}_{-0.115}$ \\
  9501-6104 & $0.423^{+0.372}_{-0.306}$ & $-0.192^{+0.693}_{-1.242}$ & $-0.774^{+1.081}_{-0.928}$ & $8.18^{+5.39}_{-4.69}$ & $218.34^{+54.69}_{-37.13}$ & $0.059^{+0.125}_{-0.127}$ \\
  9868-3704 & $0.329^{+0.283}_{-0.221}$ & $-0.507^{+0.946}_{-0.748}$ & $-0.417^{+0.724}_{-0.836}$ & $2.95^{+2.27}_{-1.58}$ & $140.97^{+30.13}_{-16.42}$ & $0.102^{+0.099}_{-0.122}$ \\
  9888-6102 & $0.478^{+0.351}_{-0.310}$ & $-0.426^{+0.891}_{-0.961}$ & $-1.131^{+0.739}_{-0.585}$ & $5.52^{+2.75}_{-2.90}$ & $201.06^{+31.32}_{-25.05}$ & $0.148^{+0.132}_{-0.099}$ \\
  \hline
 \end{tabular}
\end{table}
\end{center}

\section{Acknowledgements}
The work described in this paper was partially supported by the Seed Funding Grant (RG 68/2020-2021R) and the Dean's Research Fund of the Faculty of Liberal Arts and Social Sciences, The Education University of Hong Kong, Hong Kong Special Administrative Region, China (Project No.: FLASS/DRF 04628).





\end{document}